\def \SAIT #1 #2 {{\em Mem.\ Soc.\ Astron.\ It.\/} {\bf #1}, #2}
\def \MESS #1 #2 {{\em The Messenger\/} {\bf #1}, #2}
\def \ASTRNACH #1 #2 {{\em Astron. Nach.\/} {\bf #1}, #2}
\def \AAP #1 #2 {{\em Astron. Astrophys.\/} {\bf #1}, #2}
\def \AAL #1 #2 {{\em Astron. Astrophys. Lett.\/} {\bf #1}, L#2}
\def \AAR #1 #2 {{\em Astron. Astrophys. Rev.\/} {\bf #1}, #2}
\def \AAS #1 #2 {{\em Astron. Astrophys. Suppl. Ser.\/} {\bf #1}, #2}
\def \AJ #1 #2 {{\em Astron. J.\/} {\bf #1}, #2}
\def \ANNREV #1 #2 {{\em Ann. Rev. Astron. Astrophys.\/} {\bf #1}, #2}
\def \APJ #1 #2 {{\em Astrophys. J.\/} {\bf #1}, #2}
\def \APJL #1 #2 {{\em Astrophys. J. Lett.\/} {\bf #1}, L#2}
\def \APJS #1 #2 {{\em Astrophys. J. Suppl.\/} {\bf #1}, #2}
\def \APSS #1 #2 {{\em Astrophys. Space Sci.\/} {\bf #1}, #2}
\def \ASR #1 #2 {{\em Adv. Space Res.\/} {\bf #1}, #2}
\def \BAIC #1 #2 {{\em Bull. Astron. Inst. Czechosl.\/} {\bf #1}, #2}
\def \JSQRT #1 #2 {{\em J. Quant. Spectrosc. Radiat. Transfer\/} {\bf #1}, #2}
\def \MN #1 #2 {{\em Mon. Not. R. Astr. Soc.\/} {\bf #1}, #2}
\def \MEM #1 #2 {{\em Mem. R. Astr. Soc.\/} {\bf #1}, #2}
\def \PLR #1 #2 {{\em Phys. Lett. Rev.\/} {\bf #1}, #2}
\def \PASJ #1 #2 {{\em Publ. Astron. Soc. Japan\/} {\bf #1}, #2}
\def \PASP #1 #2 {{\em Publ. Astr. Soc. Pacific\/} {\bf #1}, #2}
\def \NAT #1 #2 {{\em Nature\/} {\bf #1}, #2}

\documentstyle{memsait}
\input epsf.sty
\begin{opening}
\title{ The VIRMOS mask manufacturing tools.\\
        (b) Mask manufacturing and handling.}
\author{ G. Conti$^1$, L. Chiappetti$^1$, E. Mattaini$^1$,
D. Maccagni$^1$, O. Le Fevre$^2$,\\
M. Saisse$^2$, G. Vettolani$^3$}
\institute{$^1$ Istituto Fisica Cosmica "G.Occhialini", CNR - Milano \\
$^2$ Laboratoire d'Astronomie Spatiale, CNRS - Marseille \\
$^3$ Istituto di Radioastronomia, CNR - Bologna }
\date{} % DO NOT INSERT ANY DATE HERE !!!
\end{opening}

\begin{document}

%\oddpagefooter{\sf Mem. S.A.It., Vol. ??, ??}{}{\thepage}
%\evenpagefooter{\thepage}{}{\sf Mem. S.A.It., Vol. ??, ??}
\oddpagefooter{}{}{} % LEAVE AS IT IS !
\evenpagefooter{}{}{} % LEAVE AS IT IS !
\ 
\bigskip

\begin{abstract}
We describe the VIRMOS Mask Manufacturing Unit (MMU) configuration, composed of two units: 
the Mask Manufacturing Machine (with its Control Unit) and the Mask Handling Unit (inclusive of Control Unit, 
Storage Cabinets and robot for loading of the Instrument Cabinets).
For both VIMOS and NIRMOS instruments, on the basis of orders received by the Mask Preparation Software 
(see paper (a) in same proceedings), the function of the MMU is to perform an off-line mask cutting and identification, 
followed by mask storing 
and subsequent filling of the Instrument Cabinets (IC). We describe the characteristics of the LPKF laser 
cutting machine and the work done to support the choice of this equipment. We also describe the remaining 
of the hardware configuration and the Mask Handling Software.
\end{abstract}

\section{INTRODUCTION}
The Mask Manufacturing Unit (MMU) is dedicated to the off-line manufacturing, 
identification and preparation of the slit masks for both VIMOS and NIRMOS instruments. The MMU includes 2 sub-units:
1) the Mask Manufacturing Machine (MMM), 
dedicated to the machining of slits in thin sheets (masks) and 2) the Mask Handling System (MHS), dedicated to the 
handling of the masks, up to the loading into the Instrument Cabinets (IC). 
Fig 1 shows the VIMOS focal plane mask reference system.
 
\begin{figure}
\centering
\epsfysize=8cm % fix the y-dimension and scales x-dim. to y-dim.
%\epsfxsize=8cm % fix the x-dimension and scales y-dim. to x-dim.
% Feel free to do the choice you prefer but do not exceed the x-dimension
% of the text lines
\hspace{5.5cm}\epsfbox{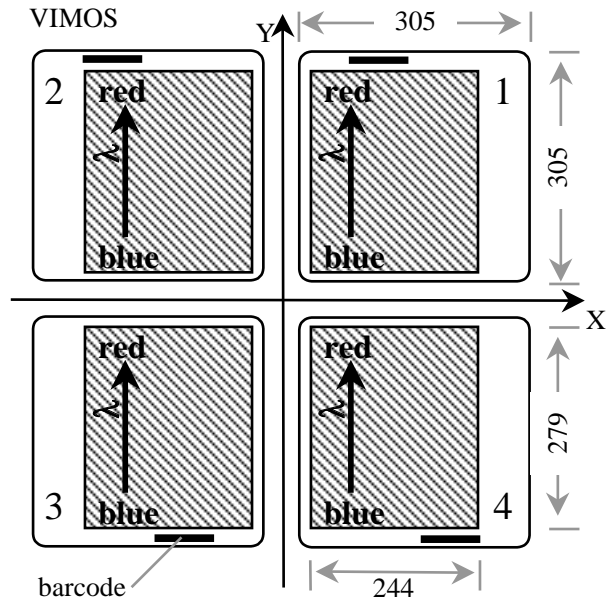} %for centering: act on hspace argument 
\caption[h]{ Mask reference system in the VIMOS focal plane. The useful areas where the slits can be located are the 
dashed ones and have dimensions of 7 $\times$ 8 arcmin. The central cross is 2 arcmin wide. The mask bar code 
is positioned in such a way that it is always on the same side when the mask are inserted in the ICs.}
\end{figure}

\section{MMU REQUIREMENTS}
\subsection{Masks}

{\bf R1)} - {\em Roughness of the slit edges.}\\
The VIRMOS Technical Specifications required $<$ 5 $\mu$m peak to peak, regardless of the slit width. 
This specification has been translated into the following quantities measurable by means of a mechanical 
roughness meter equipped with a knife-type probe:
\begin{enumerate}
\item max number of deviations from the mean $>$ $\pm$ 2.5 $\mu$m in 1 cm: 2 (parameter Pc))
\item r.m.s. as measured by the Rq parameter : $\leq$ 2 $\mu$m
\item profile shape as measured by the waviness parameter Wt : $\leq$ 3 $\mu$m.
\end{enumerate}
{\bf R2)} - {\em Time necessary to manufacture a mask.}\\
The required cutting speed is $>$ 7 m/hr. This specification is given to the MMM manufacturer as: 
$\geq$ 5 mm/sec (18 m/hr) with a quality of the cut as specified above and $\simeq$ 30 mm/sec (108 m/hr) 
without maintaining the quality of the cut needed for the slits.\\
{\bf R3)} - {\em Global slit positioning accuracy.}\\
The requirement is $<$ 30 $\mu$m, including mask positioning at the focal plane and temperature variations 
between fabrication and operation.\\ 
{\bf R4)} - {\em Unique and automatic mask identification.}\\
Each mask must be uniquely identified  for the time in which it can be used (until it is discarded).\\
{\bf R5)} - {\em Mask surfaces.}\\
The masks surfaces must have the lowest possible reflectivity at the operating wavelengths.

\subsection{Instrument Cabinets}

{\bf R6)} - Each instrument has 4 ICs (4 quadrants). Each IC can hold 15 masks and has a 
specific mechanical interface allowing it to be inserted in only one position on the instrument. The final 
design of the ICs is not yet defined at the time of writing. A remotely controlled device moves the masks 
from the ICs to the focal plane, and back. 

\subsection{Storage Cabinets}

{\bf R7)} - The manufactured mask must be temporarily stored in 2 Storage Cabinets (one for each instrument) 
waiting for the insertion in the ICs or for the discarding. Each SC is requested to contain 400 masks (100 mask-sets).

\section{CHOICE OF THE CUTTING TECHNOLOGY}

\subsection{Short history}

In the initial concept, the MMM was a milling machine which would cut the slits in a 0.1 mm thin brass 
sheet supported by an aluminium frame. We assembled a small milling machine with a 3 axes displacement 
system and a high speed mandrel (up to 80000 t/m). The minimum obtainable slit width was 300 $\mu$m. 
Because of the frames, the ICs were large and quite heavy. Furthermore, the accuracy in the slit positioning
 was hampered  by the composition of errors due to the machine positioning accuracy, the interface error
  between the mask frame and the machine working platform, and between the mask frame and the focal plane. Furthermore,
  because of 
  the thermal expansion of brass it was difficult to meet  the specification on positioning accuracy, given 
  the temperature differences between the time a mask is manufactured and used in the instrument focal plane.
Subsequent developments were aimed at minimising the sources of errors, by making use of thicker (but still 
$<$ 0.3 mm) Al unframed masks, reducing the size and the weight of the IC and eliminating the manpower 
necessary to open the frames, remove the brass sheet and put in a new one. The next natural step was the 
use of a material with a very low thermal expansion coefficient like carbon fibre, kevlar, graphite or 
Invar, but it was impossible to obtain the slit edge quality with the milling machine. 
It was only recently that a new type of laser cutting machines, called Stencil Lasers, became available 
on the market : these machines and one in particular were proven able to meet the specifications by making 
use of 0.2 mm thick Invar sheets. The cutting speed made possible to cut the mask contour on the machine 
itself, and thus it can be customised to the quadrant interface where the mask will  be placed, and, 
as a further bonus, also any slits width $>$ 100 $\mu$m became a possibility.

\subsection{Milling vs laser cutting}

For the milling technique the case of unframed aluminium masks has been considered for the comparison. 
A summary is shown here.\\
\centerline{\em Material}\\
{\bf Milling:} Aluminium (Anticorodal 100), thickness 0.3 mm, Therm.Exp.Coeff.: 23 $\mu$/m/C.\\
{\bf Laser:} Invar (Pernifer 36),  thickness 0.2 mm,  Therm.Exp.Coeff.: 0.8 $\mu$m/m/C.\\
\centerline{\em Coating}\\
{\bf Milling:} black anodization.\\	
{\bf Laser:} black antireflection paint.\\
\centerline{\em Slit characteristics}\\
{\bf Milling:} Width : from 300 to 1000 $\mu$m in steps of 100 $\mu$m. Intermediate width require 
two passes.\\
{\bf Laser:} Width : any width $>$ 100 $\mu$m.\\
\centerline{\em Slit edge quality}\\
{\bf Milling:} The requested specifications can be reached but depend strongly on the material, 
 the diameter of the cutting tool, the cutting speed,  the raw mask fixing system. The 
optimisation and the control  of these parameters is quite delicate over a long period of time. Quality 
control of the slit edge must be done very frequently.\\
{\bf Laser:} The requested specifications can be reached but depend on the manufacturers. Quality 
control can be scheduled weekly.\\
\centerline{\em Cutting of the mask contour}\\
{\bf Milling:} The mask contour cutting is difficult. Masks with pre-shaped contours are needed
 (cannot be customised).\\
{\bf Laser:} The mask contour can be cut to customise the mechanical interface of the masks to the  
focal plane assembly.\\
\centerline{\em Cutting machine components}\\
{\bf Milling:} The machine must be equipped with a high frequency mandrel, an automatic tool exchange
 mechanism, a tool checking system, a working platform with lubricating liquid tank  and a cleaning 
 unit to remove cutting oil. 
We have not found Milling Machines that completely meet our requirements on the market. A customisation 
is always necessary.\\
{\bf Laser:} Laser machines meeting our requirements can be found on the market with integrated control 
systems. \\
\centerline{\em Mask production rate}\\
{\bf Milling:} about 15 minutes for the whole cycle. This performance is the lower limit of our test 
machine and depends from the cutting tool diameter.\\
{\bf Laser:} about 10 minutes including cutting the contour of the mask.\\
%\centerline{\em Manpower needs}\\
%{\bf Milling:} The operation and  maintenance of this kind of milling machine %requires mechanical skills.\\
%{\bf Laser:} The operation and maintenance does not require specific skills.\\
%\centerline{\em Safety}\\
%{\bf Milling:} A noise absorbing cabin around the machine is necessary.\\
%{\bf Laser:} For the LPKF system the laser is classified class 1 (as CD %players).\\
%\centerline{\em Operation Cost}\\
%{\bf Milling:} The estimated cost for 2000 operation hours is from 35 to 40 %kDM.\\
%{\bf Laser:} The estimated cost for 2000 operation hours is ~ 15 kDM.\\
\centerline{\em Purchase Cost}\\
{\bf Milling:} The estimated cost of the components for a home designed milling machine plus cleaning 
unit is about 300 kDM. Engineering costs must be added.\\
{\bf Laser:} The cost of a stencil laser cutting machine ranges from 530 (Lumonics) to 630 (LPKF) kDM.

\subsection{Choice of the Laser cutting machine manufacturer}

A statement of work has been sent to 22 laser cutting machine manufacturers. 11 companies answered our enquiry, 
8 of which expressed their interest and 6 requested Invar samples to try out their product.
The most critical parameter to measure on test samples was the roughness of the slit edges. A first 
qualitative evaluation has always been done using a microscope from 50 $\times$ to 500 $\times$, to check whether\\
$-$  the slit cuts show a regular pattern (ripple) or a random noisy profile\\
$-$  the laser cutting has left some residual or re-melted material\\
$-$  the black coating has been damaged\\
$-$  the nominal slit width and shape has been respected\\ 
A quantitative evaluation of the profile roughness has been done using a mechanical roughness meter. 
The results of the measures performed on the laser cutting samples provided by the manufacturers are 
summarised in Table 1.
Tests from LPKF and Lumonics have been done with proprietary complete laser cutting machines, while 
other manufacturers have used their laser head with unspecified motion systems.

\vspace{0.5cm} %TO ALLOW SUFFICIENT SPACE BETWEEN THE TEXT AND THE FIGURES
\centerline{Table 1 - Slit edge roughness measurements}
\begin{table}[h]
\centering 
%\hspace{6.0cm} %if you want to center your table act on this argument
\begin{tabular}{|l|c|c|}
\hline
Manufacturer	&Rq = rms ($\mu$m)	&Pc = number of p-p/cm $>$ 5$\mu$m\\
\hline
LPKF						&0.8				&1\\
Lumonics (ripple-free zones)		&1.0				&3\\
Lumonics (ripple zones)			&1.8				&15\\
Rofin - Sinar				&2.6				&33\\
Haas 						&2.5				&40\\
RTM						&4.0				&30\\
{\em Required}				&{\em $\leq$ 2}		&{\em $\leq$ 2}\\
\hline
\end{tabular}
\end{table}
A high value of Pc means that the cutting edge has a residual ripple. In the table we report 2 
rows for the Lumonics tests (about 50 cutting samples in 4 successive tests with different cutting 
parameters), since we noticed that the quality of the slit edges was not constant and that some areas 
with a quasi-sinusoidal ripple were almost always present.  The effort done, together with Lumonics 
staff, to overcome the problem was not successful. This means that the Lumonics machine will need a 
tuning up of the cutting parameters to improve the performances which, so far, are critical with respect
 to the specifications.  The only machine that  completely fulfills our requirement was the  LPKF, 
 Garbsen, Germany, one.

\section{THE ADOPTED SOLUTIONS}
\subsection{Mask material}

The masks  material is Invar with thickness  0.2 mm and dimensions 305 $\times$ 305 mm. The main mechanical 
and thermal characteristics of Invar (Krupp VDM trade name is Pernifer 36), at 20 C are listed in Table 2.

\vspace{0.5cm} %TO ALLOW SUFFICIENT SPACE BETWEEN THE TEXT AND THE FIGURES
\centerline{Table 2 - Invar characteristics}
\begin{table}[h]
\centering 
%\hspace{5cm} %if you want to center your table act on this argument
\begin{tabular}{|l|l|}
\hline
chemical composition: 			&36\% Ni + 64\% Fe \\
%Curie temperature: 			&230 C \\
density:					&8.1 $g/cm^3$ \\
%thermal conductivity:			&12.8 W/mK \\
%electrical resistivity:			&76 $\mu\Omega$cm \\
modulus of elasticity:			&143 $kN/mm^2$ \\
thermal expansion coefficient:	&0.8 $10^{-6}$/C between 0 and 40 C \\
%tensile strength:				&490 N/$mm^2$ \\
\hline
\end{tabular}
\end{table}

\subsection{Mask black coating}

The mask manufacturing include, as the last operation, the cutting of the external border and thus the raw
 Invar sheets must have bigger dimensions to allow for the mechanical fixing on the working platform of the laser 
 cutting machine. A 340 by  450 mm sheet is  presently adopted. The raw masks must be:\\
$-$ coated with a black anti-reflection paint\\ 
$-$ cut to the proper size\\ 
$-$ protected against scratches\\ 
$-$ packed to be shipped to Paranal.\\ 
The requested characteristics for the coating are:\\
$-$ thickness $<$ 20 $\mu$m\\
$-$ good adhesion to the metallic substrate\\
$-$ dull black color\\
$-$ uniformity of the coating over the 2 surfaces.\\
The mask yearly need for VIRMOS is about 2400. The aim of our work was to find a method to prepare a large 
quantity of raw masks at the lowest cost.
The quotation of the 0.2 mm thick Invar  (from Krupp) is approximately 25 DM/kg for quantities of at least  1000 kg. 
A 340 $\times$ 450 Invar foil weighs 0.250 kg; from 1000 kg of material about 4000 masks can be obtained. The cost 
of the material for a single  mask is then about 6.3 DM.\\
The Invar material is delivered by Krupp in rolls with the requested width; the possible solutions to provide the
 raw masks are to cut the strip in foils and to varnish them, or to varnish the strip before cutting. We have tested 
 both possibilities. 
The second method is the cheapest one and has been tested using a strip of  stainless steel 450 mm wide. 
The process consists of:\\
$-$ chemical (alkaline bath) and mechanical (brushing) cleaning\\
$-$ coating of the two sides of the strip, using a roller system\\
$-$ warm curing of the varnish\\
$-$ insertion of a low adhesion plastic protective film\\
$-$ straightening to eliminate the roll curvature\\
$-$ cutting to the requested dimensions\\
$-$ piling of the raw masks on a transport pallet.\\ 
The performed test has produced about 600 raw masks with good results from the point of view of the quality and of 
the adhesion of the coating, but the first and the last part of the strip must be discarded and the Invar cost for 
a single mask becomes 7 DM. The cost of the whole coating process is about 6 DM per foil. So the total cost of a raw 
mask is 13 DM.

\subsection{Mask coding}
The chosen solution for mask identification was the direct cutting of a 6 digit bar code on a border of the masks using the 2/5 interleaved code. It can be read by decoders during all the operations of the Mask Handling System.

\subsection{The mask manufacturing machine}
The technical characteristics of the LPKF StencilLaser System 600 x 600 that has been choosen as the most
appropriate for our purposes, can be found on the LPKF web site www.lpkf.de/laser\_en/laser\_en.htm .\\

%\vspace{0.5cm} %TO ALLOW SUFFICIENT SPACE BETWEEN THE TEXT AND THE FIGURES
%\begin{table}[h]
%\centering 
%\centerline{\bf Table 3}
%\hspace{4.5cm} %if you want to center your table act on this argument
%\begin{tabular}{|l|l|}
%\hline
%Stand				&Natural granite\\
%Cutting area 			&600 $\times$ 600 mm\\
%XY motion 				&by DC motors\\
%Working platform 		&supported by four air pads\\
%XY accuracy (full travel )	&$\pm$ 10 $\mu$m\\
%Optical encoders resolution 	&0.5 $\mu$m\\
%Repeat accuracy			&$\pm$ 3  $\mu$m\\
%XY squareness			&$<$ 5 arcsec\\
%Accuracy of location of the	&\\
%laser cut (entire workpiece) &$\leq$ $\pm$ 15  $\mu$m\\
%Repeatability of the laser 	&\\
%cut feature dimensions   	&$\pm$ 2  $\mu$m\\
%Cutting speed range:   	&1 - 50 mm/sec\\
%Laser head: 			&Flashlamp pumped Nd:YAG (1064 nm)\\
%Max Power 				&60W\\
%Max Pulse rate 			&4000 Hz\\
%Assist gas				&air 16 bar\\
%Water cooling 			&external chiller\\
%Gas exhausting system.		&Yes\\
%Changing lamps realignment	&No\\
%Control				&PC via RS232\\
%Overall dimensions 		&1750 $\times$ 2300 $\times$ 1350 mm\\
%Dimensions of control rack  	&600 $\times$ 950 $\times$ 1900 mm\\
%Total weight 			&3000 kg\\
%Software				&CircuitCam for importing Gerber files \\ 
%					&and converting into cutting machine \\
%					&binary files;\\
%					&BoardMaster software for controlling\\
%					&the cutting machine operations.\\
%\hline
%\end{tabular}
%\end{table}

\section{THE MASK HANDLING SYSTEM}

\subsection{Current configuration}

\begin{figure}
\epsfxsize=12cm % fix the x-dimension and scales y-dim. to x-dim.
% Feel free to do the choice you prefer but do not exceed the x-dimension
% of the text lines
\hspace{5cm}\epsfbox{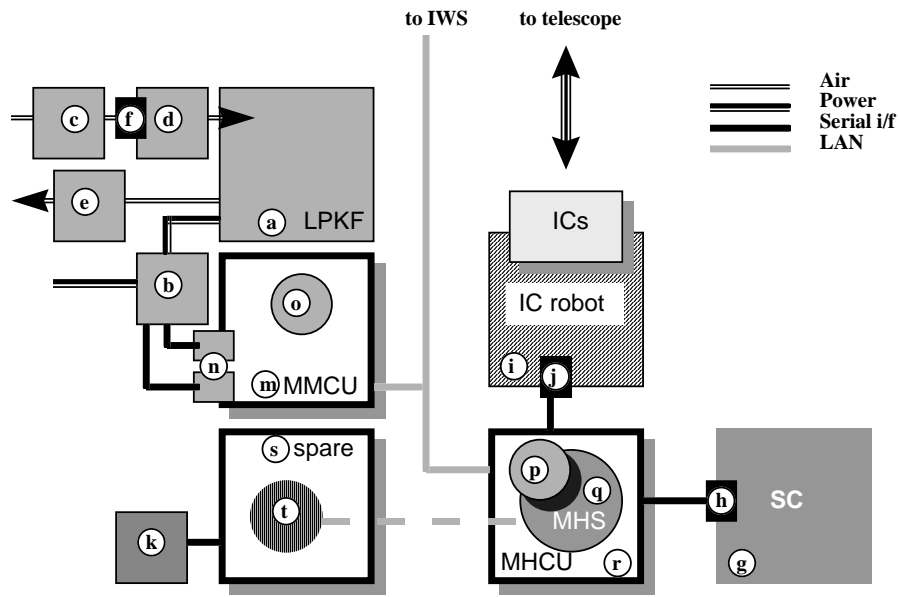} %for centering: act on hspace argument 
\caption[h]{Block diagram of the VIMOS/NIRMOS Mask Manufacturing Unit. Letters indicating the components are described in 5.1}
\end{figure}
%(a) LPKF Stencil Laser machine (b) control rack (c) cooling unit (d) pressure %duplicator (e) vacuum extractor (f) dryer/filters (g) Storage Cabinet (h) bar %code reader (i) IC robot (j) bar code reader (k) roughness meter (m) MMCU %computer (n) serial interfaces (o) BoardMaster s/w (p) CircuitCam s/w (q) MHS %s/w (r) MHCU computer (s) spare computer (t) Visual Basic development %environment}
%\end{figure}

The overall hardware and software configuration of the MMU is depicted in the block diagram in Fig.2 and includes the 
following components:\\
$-$ LPKF Stencil Laser machine model 600 x 600; ((a) in figure) with additional components supplied by LPKF\\
$-$ control electronics rack (b)\\
$-$ water cooling unit (c)\\
$-$ Maximator pressure duplicator from 8 to 16 bar (d)\\
$-$ vacuum extractor (e)\\
$-$ associated piping and cabling\\
$-$ dryer/filters (f) (additional element, not supplied by LPKF)\\
$-$ 2 serial interface cards (n) hosted in MMCU computer\\
$-$ BoardMaster software (o) (running on MMCU)\\
$-$ CircuitCam software (p)  (running on MHCU)\\
$-$ Mask Manufacturing Control Unit (MMCU) computer  (m)\\
$-$ running LPKF BoardMaster software (o)\\ 
$-$ Storage Cabinets (SC, holding 4 $\times$ 100 masks) (g), built in house, with
Datalogic DS2100 bar code reader (h) connected to serial port of MHCU\\
$-$ IC robot unit (i), under development, with
Datalogic DS2100 bar code reader (j) connected to serial port of MHCU\\
$-$ hosting Instrument Cabinets (ICs) exchanged with the instrument focal plane ; each IC has 15 numbered mask slots\\
$-$ Mask Handling Control Unit (MHCU) computer (r)\\ 
$-$ running the Mask Handling Software (q) developed in house\\ 
$-$ with slaved LPKF CircuitCam software (p)\\
$-$ bar code support software\\
$-$ Taylor - Hobson Talysurf roughness meter (k)\\
$-$ with serial connection to spare computer\\ 
$-$ Spare computer (s)\\
$-$ with roughness meter acquisition software\\
$-$ with Microsoft Visual Basic development environment (t) used for MHS.\\ 
All computers are identical Dell model Optiplex GX1 under Windows NT 4.0 Workstation with 64 Mb RAM and 
a 3$\times$2 GB disk. They are configured in an identical way (with the exception of the serial card connections), 
so that each one of them could be used as Line Replaceable Unit for all functions. In particular the spare computer 
(currently used as development environment) will be kept in cold redundancy, and only occasionally used offline with 
the roughness meter to perform quality checks on the manufactured masks.
The function of the Mask Handling Software developed in house and its interaction with the LPKF supplied software 
modules will be described below.

\subsection{Mask movement scheme}

The masks can be moved/relocated exclusively as shown in Figure 3. 
The movements between parts of the MMU system is controlled by the indicated MHS software functions 
(store, load, unload, discard). The movements inside the Instrument will be controlled by OS software functions. 
The exchange of entire ICs back and forth between Instrument and MMU buildings will be a manual operation.

\begin{figure}
\epsfxsize=12cm % fix the x-dimension and scales y-dim. to x-dim.
% Feel free to do the choice you prefer but do not exceed the x-dimension
% of the text lines
\hspace{5cm}\epsfbox{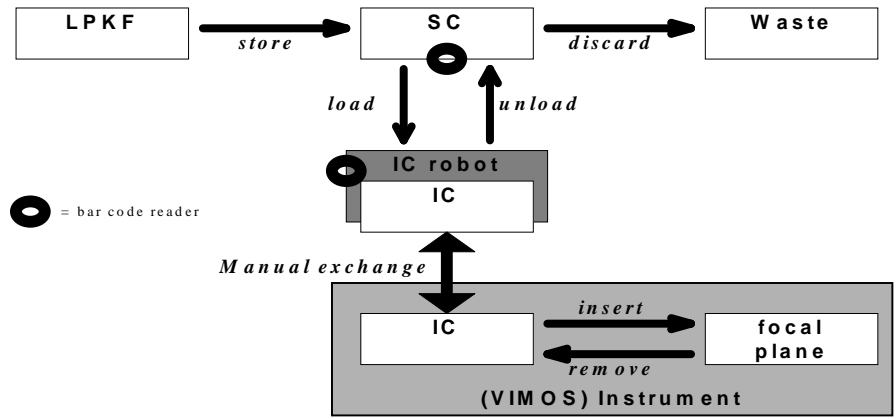} %for centering: act on hspace argument 
\caption[h]{ Scheme of allowed movements for VIRMOS masks}
\end{figure}

\subsection{Mask data files flow}

Paper (a)  describes the function of the VIMOS and NIRMOS Mask Preparation Software (MPS) as front end to 
the MHS, and outlines the concept of Orders and Reports used to regulate the flow between OHS and MPS. There
is a one-to-one correspondence between such Orders and Reports (OHS-MPS layer) with Jobs and Termination 
 reports (MPS-MHS layer). For each Order sent by OHS to MPS, MPS  sends a Job to MHS. MHS sends a Termination 
 report to MPS, which uses it to generate a Report to OHS.\\
Note that MPS is responsible to associate to each Observing Block (OB) a mask set, i.e. 4 masks identified by a unique 
5 digit identifier. The 6-digit barcode is composed prepending a 1-digit quadrant identifier (1-4 for VIMOS and 
5-8 for NIRMOS) to the mask code. MHS only knows about masks (sets), and knows nothing about OBs.\\
A Job is an ASCII file with a list of mask identifiers, and a Termination report is a similar ASCII file
 associating  an array of status codes to each mask. In addition to Jobs and Termination reports, there are other 
 types of files exchanged between MPS and MHS.\\
$-$ A series of Machine Slit Files (MSFs) generated by MPS and associated to a Mask Manufacturing Job. 
They are described in paper (a)\\
$-$Storage Cabinet Table files (SCT) are maintained by MHS and list all the masks currently stored in SCs.\\
$-$Instrument Cabinet Table files (ICT) are maintained by MHS and list which masks are currently loaded in the 
numbered slot of each IC.\\
The exchange of files between MPS and MHS  occurs, for security reasons, exclusively via ftp sessions 
initiated by the MPS side. MPS will put Jobs (and eventual MSFs) into an instrument staging area on MHCU.
MHS will move the files being processed and place back Termination reports and a copy of the ICT and SCT in the same area, from which MPS will get them.
Further file types are used internally in MHS as described below. 

\subsection{Description of the MMU cycles}

In the following we describe the procedures used during the typical lifetime of a mask set required for spectroscopic 
observations. Different (simplified) procedures may apply to masks required for instrument maintenance.

\subsubsection{Manufacturing and storage cycle}

$-$ In response to a Mask Manufacturing Order sent from OHS to MPS, MPS translates it  into a Mask Manufacturing 
Job (MMJ) for MHS, and supplies an ASCII Machine Slit File (MSFs) for each mask\\
$-$ MHS convert function in turn:\\
$--$ converts all MSFs to the CAD industrial standard Gerber format\\
$--$ runs LPKF CircuitCam to convert Gerber files into proprietary binary format (LMD)\\
$--$ moves LMD files for entire mask sets to the MMCU disk.\\
$-$ Only complete mask sets (all four quadrant successfully converted) are considered for manufacturing.\\
$-$ Operator (on MMCU) uses the LPKF BoardMaster program to manufacture one mask at a time.\\
$-$ Masks are manufactured and stored 4 by 4 into an intermediate repository, to prevent storage of incomplete mask sets.\\
$-$ Operator (on MHCU) uses MHS store function to identify and store all masks of a mask set in the Storage Cabinet. 
MHS store function updates the SCT (Storage Cabinet Table) and generates a Mask Manufacturing Termination report (MMT) for MPS.\\
$-$ MPS translates the MMT into a Mask Manufacturing Report for OHS.

\subsubsection{ Loading and unloading cycle}

$-$ Some time later (at least one night in advance of the observation) OHS issues a Mask Insertion Order to MPS, 
and MPS translates it into a Mask Insertion Job (MIJ) for MHS.\\
$-$ Instrument Cabinets (ICs) are physically moved from instrument to IC robot\\ 
$-$ Operator on MHCU runs MHS unload function, which arranges for any mask not in MIJ to be unloaded from IC 
and put back in SC, while leaves in IC any mask already there and requested by MIJ.\\
$-$ Operator on MHCU runs MHS load function, which arranges to load from SC into IC any mask in MIJ not already
 loaded and generates a Mask Insertion Termination report (MIT) for MPS. Both functions also update ICT and SCT 
 as appropriate\\
$-$ MPS translates the MIT into a Mask Insertion Report for OHS.

\subsubsection{Discarding cycle}
 
$-$ Some time later OHS issues a Mask Discard Order to MPS for OBs which have either been successfully executed 
or have expired and MPS translates it  into a Mask Discarding Job (MDJ) for MHS.\\
$-$ Operator on MHCU runs MHS discard function, which arranges for masks to be removed from SC and associated 
LMD files to be deleted from MMCU (where they have been kept until this time to allow reproduction in case of 
damages : this means that such masks cannot be manufactured again any longer) and generates a Mask Discard 
Termination (MDT) report for MPS.\\
$-$ MPS translates the MIT into a Mask Discard Report for OHS: as a result the relevant OBs are marked as no longer 
schedulable.

\end{document}